\documentclass[seceq]{ptptex}

\usepackage{graphicx}
\usepackage{wrapft}
\usepackage{hyperref}

\def\bfm#1{\mbox{\boldmath $#1$}}

\newcommand{\MeV}{\,{\rm MeV}}
\newcommand{\ie}{{i.e.}}
\newcommand{\beq}{\begin{equation}}
\newcommand{\eeq}{\end{equation}}
\newcommand{\ba}{\begin{array}}
\newcommand{\ea}{\end{array}}
\newcommand{\diag}{{\rm diag}}
\newcommand{\Sqrt}[1]{\sqrt{\mathstrut #1}}
\newcommand{\ds}[1]{
  \setbox0=\hbox{\ensuremath{#1}}
  \hbox to\wd0{\hbox to0pt{\hbox to\wd0{\hss/\hss}\hss}\box0}}

\title{%
Polyakov-Nambu-Jona Lasinio model \\
and Color-Flavor-Locked phase of QCD
}

\author{%
Hiroaki \textsc{Abuki}%
}

\inst{%
I.N.F.N., Sezione di Bari, I-70126 Bari, Italia\footnote{
moved to Frankfurt Institute for Advanced Studies (FIAS), 
Johann Wolfgang Goethe-Universit{\"a}t, D-60438 Frankfurt am Main, Germany; 
email:~abuki@th.physik.uni-frankfurt.de}
}

\abst{%
The effect of Polyakov loop on the QCD phase diagram at high density
is studied within the Nambu-Jona Lasinio model with Polyakov loop
(PNJL model).
We point out that the color neutrality is missing in the standard PNJL
model at finite density.
Moreover, we discuss how the color-flavor locked (CFL) phase is to be
distorted by the inclusion of Polyakov loop. 
}

\begin{document}
\maketitle

\section{Introduction}
The phase diagram of strongly interacting matter 
has been a subject of theoretical/experimental work since the
foundation of Quantum chromodynamics (QCD).
The perturbative QCD can be of some help in the extremely high density
regime, but it is no longer reliable at density of physical interest.
The lattice QCD is a powerful tool to study such a strongly interacting
regimes.
However there is a well-known difficulty in simulating QCD on lattice at
finite density.
So exploring phase structure at intermediate density, where neither 
lattice simulations nor perturbative calculations can be trusted,
remains the subject of various model studies which mimic some of basic
features in QCD. 
The Nambu-Jona Lasinio (NJL) model is one of them
\cite{Nambu:1961tp}
and it nicely predicts the chiral restoration of QCD at extreme
conditions \cite{Hatsuda:1994pi}.

The main defect of the NJL model had been the lack of the notion of the
confinement. 
To improve this point, Fukushima included the Polyakov loop dynamics
into the NJL model \cite{Fukushima:2003fw}, and the model is now called
``Polyakov-Nambu-Jona Lasinio'' (PNJL) model.
This model has two order parameters, $q\bar{q}$ for the chiral restoration,
and the Polyakov loop $\Phi$ for the deconfinement.
Even though these two serve as the exact order parameters only in
the different limits, ($m_q\to 0$ and $m_q\to\infty$), 
the model enables to interpret nicely some bulk properties of matter
observed on the lattice on the field theoretical ground
\cite{Ratti:2005jh}.

In this work, we will extend the application of PNJL model to color
superconducting phases at high density \citen{Abuki:2008ht}.
In particular, we are interested in: 
(a)~how the phase structure in $(T,m_s^2/\mu)$-plane will be modified by
the inclusion of the Polyakov loop, and
(b)~what is the consequence of imposing the color/electrical neutralities
on the PNJL model with and without diquark condensations.
The purpose (a) is regarded as the extension of the earlier work
\citen{Fukushima:2004zq,Abuki:2004zk}, while (b) is
considered as the extension of
Refs.~\citen{Roessner:2006xn,Ciminale:2007ei}.
We note there is a parallel development on the role of electrical
neutrality in the PNJL model at low density \cite{Abuki:2008tx}.

\section{The model}\label{sec:model}
The Wilson line operator is a key quantity whose expectation value
plays a role of order parameter for deconfinement transition
in a pure gauge theory.
It can be expressed by the background Euclidean
temporal gauge field
$A_4(\tau,\bfm{x})\equiv igA_0^\alpha(\tau,\bfm{x})%
T_\alpha$ ($T_\alpha=\frac{\lambda_\alpha}{2}$; $\{\lambda_\alpha\}$ are
the standard Gell-Mann matrices for SU$(3)_{\rm c}$) as
\beq
\ba{rcl}
  L_Q=P_\tau\exp\left(i\int_0^{1/T}d\tau
  A_4(\tau,\bfm{x})\right).
\ea
\eeq
The Wilson line in the anti-triplet representation can be defined as
$L_{\bar{Q}}\equiv L_Q^\dagger$.
In a pure gauge theory with zero chemical potential for quarks
$(\mu=0)$, 
these are regarded as the operators associated with heavy (anti)quark
excitation in the gluonic heat bath at temperature $T$; loop
$\Phi=\frac{1}{N_c}\langle{\rm tr}L_Q\rangle_{T}$
(anti-triplet loop $\bar{\Phi}=\frac{1}{N_c}\langle{\rm
tr}L_{\bar{Q}}\rangle_{T}$)
is related to the Free energy of single (anti)quark excitation in the
gluon medium by $\Phi=e^{-F_{Q}/T}$
and $\bar{\Phi}=e^{-F_{\bar{Q}}/T}$.
In this case, one finds eventually $\bar{\Phi}=\Phi=\Phi^*$.
In the full gauge theory with dynamical quarks and with a finite
chemical potential for quarks, however, this is no longer the case
because quark and antiquark propagate differently in each direction
of imaginary time.
The detailed analysis of a matrix model shows that $\Phi$ and
$\bar{\Phi}$ 
certainly differ from each other but still both stay
real; this is due to the imaginary piece in the action proportional to
the imaginary part of $\frac{1}{N_c}{\rm tr}L_Q$ which comes from
integration of dynamical quarks and is $C$-odd quantity
\cite{Dumitru:2005ng}.

In the PNJL model, the loop $\Phi$ can be parametrized by eight {\em real}
parameters $\{\varphi_1,\varphi_2,\cdots,\varphi_8\}$ each of which has
a dimension of energy, as
\beq
\ba{rcl}
  \Phi[A_4]=\frac{1}{N_c}{\rm tr}e^{iA_4/T},\quad %
  A_4=\sum_{\alpha=1}^{N_c^2-1}\varphi_\alpha T_\alpha.
\ea
\eeq
Using this $A_4$ field, the PNJL model is given by the following
Lagrangian density
\beq
 {\mathcal L}[q,\bar{q}; A_4]=\bar{q}(i(\ds{\mathcal D}[A_4]
  +\gamma_0(\mu+\delta\mu_{\rm eff}))q%
  +\textstyle\frac{G}{4}\bar{q}P_{\eta\eta'}\bar{q}^Tq^T\bar{P}_{\eta\eta'}q%
  -{\mathcal U}(T,\Phi[A_4]).
\label{eq:ourmodel}
\eeq
$q=(q_{ur},q_{dg},q_{sb},\,q_{ug},q_{dr},\,q_{sr},q_{ub},\,q_{db},q_{sg})^T$
is the quark field.
${\mathcal D}_\mu=\partial_\mu-\delta_{\mu 0}A_4$
is the covariant
derivative through which the Polyakov loop can change the nature of 
propagation of dynamical quarks.
$G$
parametrizes the strength of attractive coupling in the color-flavor
channel $P_{\eta\eta'}=C\gamma_5\epsilon_{\eta ij}\epsilon_{\eta'ab}$
($\bar{P}_{\eta\eta'}=\gamma_0 P_{\eta\eta'}^\dagger\gamma_0$). 
In this work, we take the CFL type {\em diagonal} ansatz for diquark
condensation \cite{Alford:1998mk}, \ie,
\beq
\ba{rcl}
 \frac{G}{2}\langle q^T \bar{P}_{\eta\eta'}q\rangle=\left(%
 \begin{array}{ccc}
  \Delta_1 & 0 & 0 \\
  0 & \Delta_2 &0 \\
  0 & 0 & \Delta_3\\
  \end{array}
 \right)_{\eta\eta'}\equiv\bfm{\hat{\Delta}}_{\eta\eta'}.
\ea
\label{condensate}
\eeq
$\eta$ ($\eta'$) stands for the flavor (color) index.
We work within the chiral SU$(2)$ limit setting $m_u=m_d=0$, 
and take into account the strange quark mass $m_s$ within the high
density approximation. So we set
$\delta\mu_{\rm eff}=-\mu_eQ+\mu_3 T_{3}+\mu_8 T_{8}%
-\frac{m_s^2}{2\mu}\diag.(0,0,1)_{\rm f}\times{\bf 1}_{\rm c}$,
where $Q=\diag.(2/3,-1/3,-1/3)_{\rm f}\times{\bf 1}_{\rm c}$,
$T_3={\bf 1}_{\rm f}\times\frac{1}{2}\lambda_3$, 
and $T_8={\bf 1}_{\rm f}\times\frac{1}{\Sqrt{3}}\lambda_8$.
$\mathcal U(T,\Phi)$ 
is the Polyakov loop potential which controls the
confinement/deconfinement transition in the pure gauge sector,
whose detailed form will be given later.

The real part of effective potential\footnote{We have also an imaginary
part of the effective potential, which may viewed as a sign problem. 
It can cause a splitting of $\Phi$ and $\bar{\Phi}$, but here we simply
discard it \cite{Roessner:2006xn}.
Accordingly, we have $\Phi=\bar{\Phi}$.} 
within the high density effective theory (HDET) comes out to be
\beq
\ba{rcl}
 \Re\Omega(\Delta_\eta,\varphi_\alpha)%
 &=&{\mathcal U}(T,\Phi)%
  -\frac{\mu_e^4}{12\pi^2}-\frac{\mu_e^2T^2}{6}-\frac{7\pi^2 T^4}{180}\\[1ex]
 & & -\sum_{A=1}^{9}\left[\frac{(\mu+\delta\mu_{\rm eff}^{A})^4}%
  {12\pi^2}+\int\frac{(\mu+l_\parallel)^2%
  dl_{\parallel}}{2\pi^2}%
  2T\ln(|\!|1+e^{-E_A(l_\parallel)/T}|\!|)\right]\\[1ex]
 & &+\sum_\eta\frac{\Delta_\eta^2}{G}%
 -\sum_{A=1}^9
  \int_{-\omega_c}^{\omega_c}\frac{(\mu+l_\parallel)^2%
  dl_{\parallel}}{2\pi^2}%
  \Big[\Re E_A(l_\parallel)%
  -|l_{\parallel}-\delta\mu_{\rm eff}^{A}|\Big].\\[1ex]
\ea 
\eeq
$l_\parallel$
is the quark momentum measured from the Fermi surface
$p=\mu$.
$\omega_c$
is the ultra-violet cutoff needed to regularize the divergent third line.
The complex energies $\{E_1,E_2,\cdots,E_9\}$ are defined by choosing
the eigenvalues of non-hermitian matrix
\beq
\ba{rcl}
 {\mathcal H}%
 =\left(\begin{array}{cc}
   l_\parallel-\delta\mu_{\rm
    eff}+iA_4&\Delta_\eta\epsilon_{\eta ab}\epsilon_{\eta ij}\cr
   \Delta_\eta\epsilon_{\eta ab}\epsilon_{\eta ij}%
   &-l_\parallel+\delta\mu_{\rm eff}^t-iA_4^t\cr
   \end{array}
  \right),
\ea
\eeq 
such that $E_A\to |l_\parallel-\mu_{\rm eff}^A|$ when $\Delta_\eta\to0$
is satisfied.
We use the cutoff ($\omega_c$) dependent coupling
$\frac{1}{G}=\frac{2\mu^2}{\pi^2}%
\ln\left(\frac{2\omega_c}{2^{1/3}\Delta_0}\right)$ where $\Delta_0$ is
the magnitude of CFL gap parameter at $T=0$ in the chiral limit
($m_s=0$).
With this convention, the effective potential is only weakly
(logarithmically) divergent, so the gap equations and neutrality
conditions have well-defined finite limit as $\omega_c\to\infty$.

\vspace*{1mm}
\noindent
{\bf Parameter reduction via gauge invariance:~}
The effective potential is a function of three gap parameters
$\{\Delta_\eta\}$ 
and eight parameters $\{\varphi_\alpha\}$ for the Wilson line matrix.
In certain cases, the gauge invariance is helpful to reduce the
dynamical variables.
$\Delta_\eta=0$ 
is such a case; the number of parameters for $A_4$ 
can be reduced from $N_c^2-1$ down to $N_c-1$
as we see below. 
Since the Wilson line transforms as $L_Q\to gL_Qg^{-1}$ by the gauge
transformation with $g$ being the arbitrary SU$(3)$ matrix, 
we see
\beq
\ba{rcl}
 \Omega(A_4)=\Omega(gA_4g^{-1}),
\ea
\eeq
from the gauge invariance.
We can always make $A_4$ diagonal by choosing the suitable $g$
as $gA_4g^{-1}=\phi_3 T_3 +\phi_8 T_8$, so we can work with this
simplified ansatz for $A_4$ without loss of generality. 
The effective potential as well as the Polyakov loop $\Phi$
is a function of $\{\phi_3,\phi_8\}$ in this case. 
This procedure is just like the change of integration variable from
SU$(3)$
to its eigenvalues by integrating out six phase variables.
However, once diquarks come into the problem, this simple
reduction does no longer work.
In fact, if we try to diagonalize $A_4$, the diquark condensate also
suffers from the gauge rotation
\beq
\ba{rcl}
  A_4\to gA_4g^{-1},\quad%
  \bfm{\hat{\Delta}}_{\eta\eta'}
  \to\bfm{\hat{\Delta}'}_{\eta\eta'}\equiv%
  \left(\bfm{\hat{\Delta}}g^{-1}\right)_{\eta\eta'}.
\ea
\eeq
We note that the new condensate matrix $\Delta'_{\eta\eta'}$ is no longer
restricted to be of diagonal form in the color-flavor space.
$A_4$ 
can be diagonalized by $g$ which can be parametrized by six ``phase''
parameters; at the same time the diquark condensate acquires this phase
rotation, so the new diquark condensate is to be parametrized by nine
parameters.
Thus in principle, we have two choices; one is (1) to work in the diagonal
form of $A_4$ with the generalized off-diagonal ansatz for diquark
condensate. 
The other is (2) to work in the standard diquark ansatz with
$\Delta_\eta$ 
but with a general $A_4$ parametrized by eight parameters.
These proper treatments require us to deal with all eleven variational
parameters.
Leaving these proper arguments to our future plan, we here work within
the simplified {\em ansatz} for the ground state that the diquark
condensate is diagonal $\{\Delta_\eta\}$ even after fixing the gauge
which diagonalizes $A_4$;
this means we take only the diagonal entries of $A_4$,
$\{\varphi_3,\varphi_8\}$, 
as the variational variables.

Although the continuous gauge freedom should be considered 
to have gone away to diagonalize $A_4$, 
there remain six discrete gauge transformations each of which leaves
$\Phi$ 
unchanged; these are the elements of permutation of fundamental color
indices.
Also in accordance with discarding $\Im\Omega$, we further put $\Phi$ to be
real. For this, we take $\varphi_8=0$, so in this way, the gauge is
completely fixed in our calculations.

Finally, we specify the Polyakov loop potential.
We adopt the following form
\citen{Ratti:2005jh}
\beq
\ba{rcl}
\frac{{\mathcal U}(T,\Phi)}{T^4}=-\frac{b_2(T)}{2}\Phi^*\Phi%
  +b(T)\log\left(1-6\Phi^*\Phi+4(\Phi^{*3}+\Phi^3)-3(\Phi^*\Phi)^2\right),
\ea
\eeq
with 
$b_2(T)=a_0+a_1\left(\frac{T_0}{T}\right)+a_2\left(\frac{T_0}{T}\right)^2$, 
$b(T)=b_3\left(\frac{T_0}{T}\right)^3$.
$T_0$
is the value of the transition temperature for deconfinement in
pure gauge, \ie, $T_0=270\MeV$. 
The logarithmic term was first proposed in \citen{Fukushima:2003fw} and
it is nothing but the Vandermonde determinant, \ie, the Jacobian
associated with the
change of dynamical variables from the SU$(3)$ matrix to its eigenvalues
\citen{Kogut:1981ez}.
For more details for parameter setting used in our numerical analysis,
the readers are referred to Ref.~\citen{Abuki:2008ht}.

\section{Results and Discussion}
\begin{figure}[tp]
\centerline{%
  \includegraphics[width=0.41\textwidth,clip]{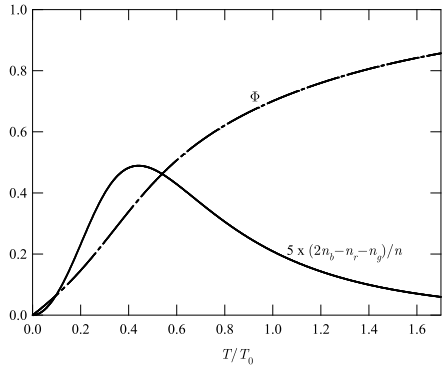}
}
 \caption[]{$T_8$ color density with $m_s=\Delta_\eta=\mu_{e,3,8}=0$
 as a function of $T$.
}
 \label{cdensity}
\end{figure}
\noindent
{\bf Color density associated with color symmetry breaking:~}
Before going into the full calculation, we study the simplified case
with unpaired matter in the chiral limit, $\Delta_\eta=m_s=0$,
just to illustrate the importance of imposing color neutrality in the
PNJL model at finite $\mu$.
In Fig.~\ref{cdensity}, we show the color $T_8$ density, 
$\langle q^\dagger T_8q\rangle$, 
and the Polyakov loop $\Phi$ as a function of $T$.
Surprisingly, $T_8$ color density takes nonzero value.\footnote{%
It should be noted that the color density itself is a gauge
dependent quantity and thus should depend on the choice of the
gauge. 
With our diagonal representation of $A_4$ with $\varphi_8=0$, the $T_8$
color density becomes finite as we observed above. 
If we selected a different gauge, the other entries of octet color
density $\{\langle q^\dagger T_\alpha q\rangle\}$ should have appeared.
The important thing is, however, whichever gauge we
choose, some color density should become finite; in fact the
squared sum of the octet color densities is shown to be the gauge
independent quantity \citen{Buballa:2005bv}.}
It can be shown that this is the case except for two different limits,
$T\to0$ 
and $T\to\infty$ $(\Phi\to 1)$ \cite{Abuki:2008ht}.
The Polyakov loop $\Phi$ is the colorless object, so one might think
it is strange to have nonvanishing color density.
The reason is simple; we are breaking color symmetry in addition to
$Z_3$ 
center symmetry by introducing the constant $A_4$ background field.
In fact, $\Phi=\frac{1}{N_c}{\rm tr}L_Q$ is invariant under the
color rotation $L_Q\to gL_Qg^{-1}$; it changes its value only under
color transformation which is not exactly periodic in imaginary time
but only up to $Z_3$, \ie, $\Phi\to z\Phi$ under
\beq
\ba{rcl}
L_Q\to gL_Q(zg)^{-1},\quad\mbox{where}\quad z\in Z_3.
\ea
\eeq
Since we {\em assumed} the constant $A_4$ background to parametrize
$\Phi$, 
and $A_4$ in contrast to $\Phi$ is not invariant under color rotation,
we have broken the color symmetry in addition to $Z_3$ symmetry.
This is unexpected, undesirable feature of the PNJL model and may be
considered as the model artifact. It could be dangerous for theoretical
foundation of the model itself, but we do not discuss further this
problem here.
Instead, we simply assume that the model is still useful once we
impose the vanishing color density as the constraint by tuning color
chemical potentials $\{\mu_3,\mu_8\}$.\footnote{%
We checked that off-diagonal color charge densities 
$\langle q^\dagger T_\alpha q\rangle$
(for $a\ne3,\,8$) are automatically vanishing
for all the situations we are interested in here.}

\begin{figure}[tp]
\centerline{%
  \includegraphics[width=0.41\textwidth,clip]{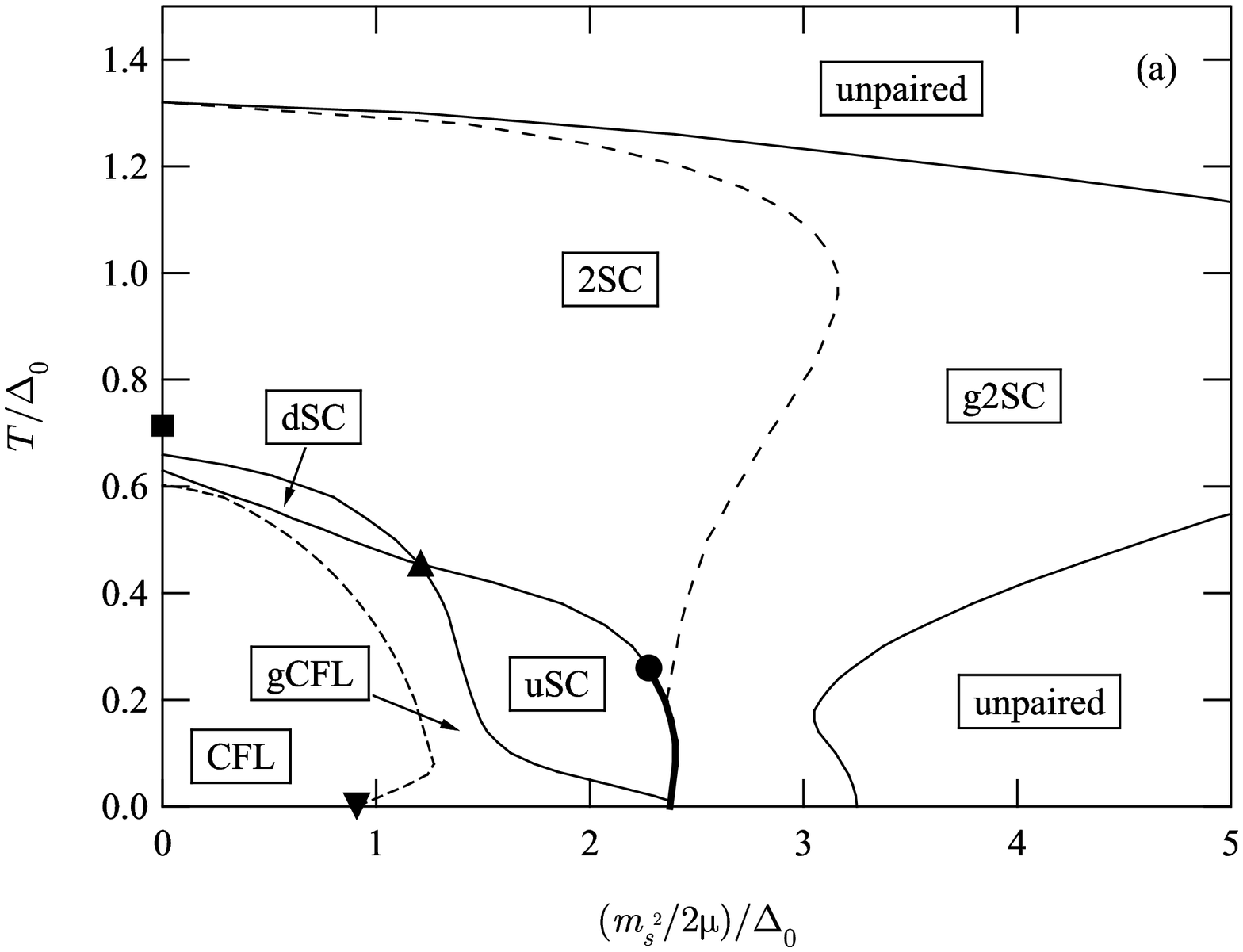}
  \includegraphics[width=0.41\textwidth,clip]{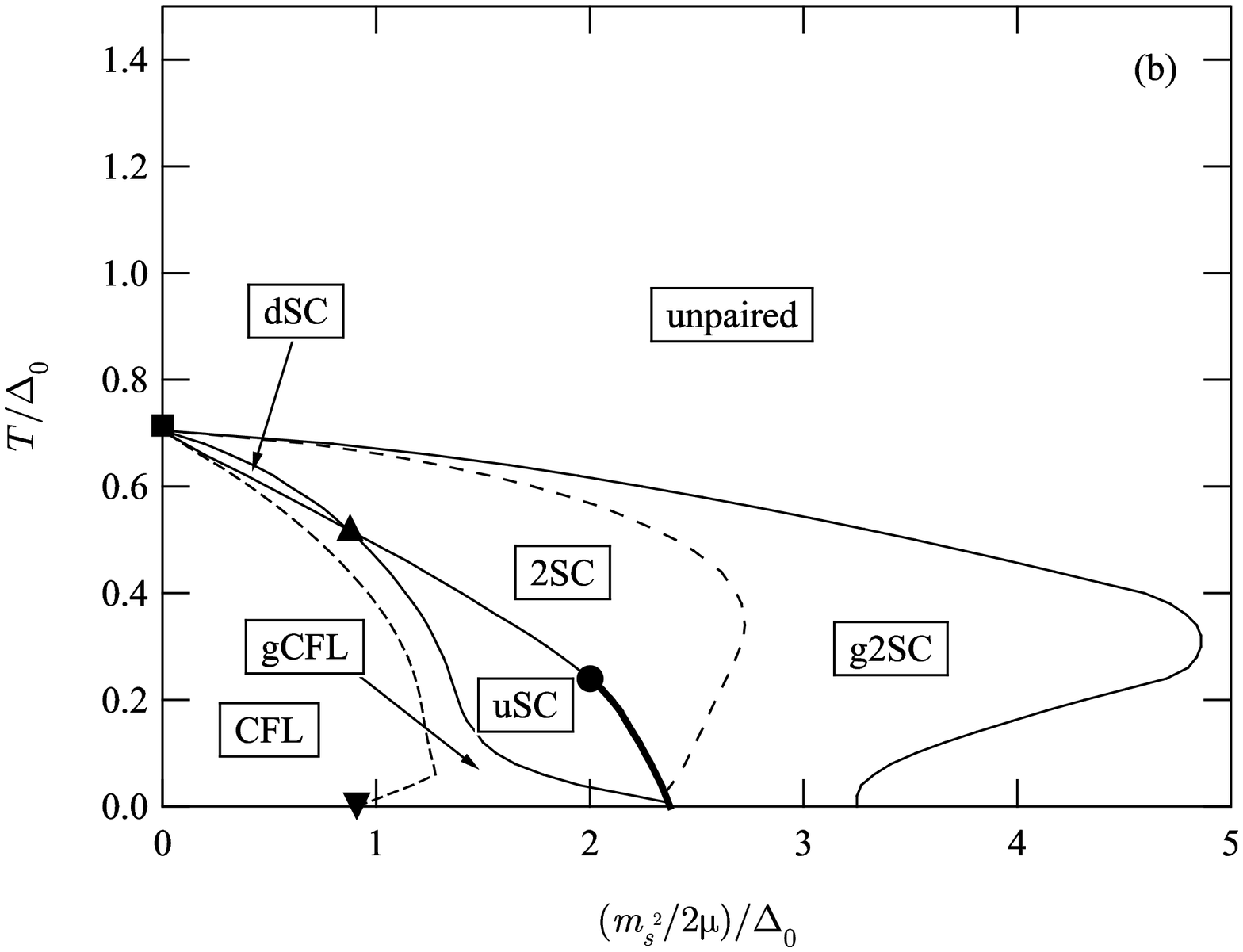}%
}
 \caption[]{
 {\scriptsize (Reprinted figure with permission from [%
 \href{http://link.aps.org/abstract/PRD/v77/e074018}%
 {H.~Abuki {\em et al.}, Phys.\ Rev.\  D {\bf 77}, 074018 (2008)}]
 Copyright (2008) by the American Physical Society).}
 (a):~Phase diagram in $(\frac{m_s^2}{2\mu},T)$-plane
 at $\Delta_0=60\MeV,\,\mu=500\MeV$. 
 (b):~The same as (a) but without the Polyakov loop.
 }
 \label{phase}
\end{figure}

\vspace*{1mm}
\noindent
{\bf The phase diagrams:~}
The phase diagram coming out from our high density PNJL model is
displayed in Fig.~\ref{phase}. (a).
For comparison, we have also shown in Fig.~\ref{phase}. (b), the phase
diagram calculated with the model without the Polyakov loop.
From these figures, the impact of the Polyakov loop dynamics on the
quark Cooper pairing is clear; it has two major effects.

\vspace*{1mm}
\noindent

(a) First, we notice that the Polyakov loop dynamics stabilizes the 2SC
phase significantly.
In fact, the critical temperature for the 2SC-to-unpaired phase
transition at $m_s=0$ is almost doubled by inclusion of the
Polyakov-loop dynamics.
This point can be understood by the observation that the Polyakov loop
suppresses the thermal excitation of colored quasiquarks which tend to
break the Cooper pair condensate \cite{Abuki:2008ht}.
Numerically, the factor of enhancement of $T_c$ is $1.8$ which is
in good agreement with our analytical estimate $1.79$
\cite{Abuki:2008ht}.

\vspace*{1mm}
\noindent
(b) Second, as a consequence of the effect (a), we have the color-flavor
unlocking transition even at $m_s=0$.
One may wonder why the SU$(3)$ flavor symmetry should be broken down to
the {\em isospin} SU$(2)$ in the $(u,d)$-sector,
and why not either in $(s,u)$ or $(d,s)$ sector. 
This is strange because at $m_s=0$ the flavor SU$(3)$ symmetry is
perfect so how can the flavor-blind Wilson line distinguish them? 
Actually, the fact that we have the isospin symmetry intact is
directly attributed to our model assumption mentioned in
Sec.~\ref{sec:model};
we are limiting ourselves to treat only two out of eight parameters for
the SU$(3)_{\rm c}$ matrix, $L_Q$. 
As noted, in principle this can not be justified in our case because
the gauge is already fixed in the diquark sector once we put the
ansatz for diquarks to the diagonal form 
$\hat{\Delta}_{\eta\eta'}=\diag.(\Delta_1,\Delta_2,\Delta_3)$; 
therefore there remains no
continuous gauge freedom to rotate $L_Q$ to a diagonal form.
So this limitation should be rather viewed as {\em an ansatz} for the
many possible ground states, such as the color-flavor locking ansatz.
For the proper treatment, we should take into account all the eight
parameters to represent the Wilson line $L_Q$.
It is possible after making such a proper treatment,
either that the ground state prefers the diagonal form of $L_Q$
or that the ground state we obtained here turns out to be one of several
degenerated ground states.
We defer this task in future.

\section*{Acknowledgements}
The author thanks R. Gatto, G. Nardulli, and M. Ruggieri for the
fruitful collaboration.
He also thanks T. Kunihiro and the other members of
organizing committee for the YITP symposium on ``Fundamental Problems in
Hot and/or Dense QCD'' for their kind invitation.


\end{document}